# Direct observation of band structure modifications from monolayer WSe₂ to Janus WSSe

*Masato Sakano[1][*][†], Shunsuke Akatsuka[1], Takato Yamamoto[1], Tianyishan Sun[2,3], Dingkun Bi[2,3], Hiroto Ogura[2,3], Naoya Yamaguchi[4], Fumiyuki Ishii[4], Natsuki Mitsuishi[5][‡], Kenji Watanabe[6], Takashi Taniguchi[7], Miho Kitamura[8,9], Koji Horiba[9], Kenichi Ozawa[8], Katsuaki Sugawara[10, 11, 12], Seigo Souma[11,13], Takafumi Sato[10,11,13,14,15], Yuta Seo[16], Satoru Masubuchi[16], Tomoki Machida[16], Toshiaki Kato[2,3], and Kyoko Ishizaka[1, 5]*

AUTHOR ADDRESS

[1]Quantum-Phase Electronics Center and Department of Applied Physics, The University of Tokyo, Bunkyo-ku, Tokyo, 113-8656, Japan
[2]Department of Electronic Engineering, Tohoku University 980–8579 Sendai, Japan
[3]Advanced Institute for Materials Research (AIMR), Tohoku University Sendai 980–8577, Japan
[4]Nanomaterials Research Institute (NanoMaRi), Kanazawa University, Kakuma-machi, Kanazawa 920-1192, Japan
[5]RIKEN Center for Emergent Matter Science (CEMS), Wako, Saitama, 351-0198, Japan
[6]Research Center for Electronic and Optical Materials, National Institute for Materials Science, 1-1 Namiki, Tsukuba 305-0044, Japan
[7]Research Center for Materials Nanoarchitectonics, National Institute for Materials Science, 1-1 Namiki, Tsukuba 305-0044, Japan
[8]Institute of Materials Structure Science, High energy Accelerator Research Organization (KEK), Tsukuba 305-0801, Japan
[9]Institute for Advanced Synchrotron Light Source, National Institute for Quantum Science and Technology (QST), Sendai 980-8579, Japan
[10]Department of Physics, Graduate School of Science, Tohoku University, Sendai 980-8578, Japan
[11]Advanced Institute for Materials Research (WPI-AIMR), Tohoku University, Sendai 980-8577, Japan
[12]Precursory Research for Embryonic Science and Technology (PRESTO), Japan Science and Technology Agency (JST), Tokyo 102-0076, Japan
[13]Center for Science and Innovation in Spintronics (CSIS), Tohoku University, Sendai 980-8577, Japan
[14]International Center for Synchrotron Radiation Innovation Smart (SRIS), Tohoku University, Sendai 980-8577, Japan
[15]Mathematical Science Center for Co-creative Society (MathCCS), Tohoku University, Sendai 980-8578, Japan
[16]Institute of Industrial Science, The University of Tokyo, Meguro-ku, Tokyo 153-8505, Japan

ABSTRACT

Janus monolayer transition metal dichalcogenides (TMDs), created by post-growth substitution of the top chalcogen layer, represent a new direction for engineering 2D crystal properties. However, their rapid ambient degradation and the difficulty of obtaining large-area monolayer samples have limited the available experimental probes, leaving their detailed electronic structure near the Fermi level largely unexplored. In this work, by performing micro-focused angle-resolved photoemission spectroscopy (μ-ARPES) on an identical sample transformed from monolayer $WSe_2$ to Janus WSSe via a $H_2$ plasma-assisted chalcogen-exchange method, we reveal the evolution of its electronic band structure. We observe ARPES signature consistent with the Rashba-type spin splitting due to broken horizontal mirror symmetry, and a significant upward shift of the highest valence band at the Γ–point by approximately 160 meV. These direct observations clarify the key electronic modifications that govern the material's properties and provide a pathway for band engineering in Janus TMDs.

MAINTEXT

The advent of two-dimensional (2D) crystals—including graphene,[1] hexagonal boron nitride (h-BN),[2,3] transition-metal dichalcogenides (TMDs),[4,5] black phosphorus[6]—and their assembly into van der Waals (vdW) heterostructures[7] has provided an unprecedented platform for creating materials with novel functionalities. Prior research in this field has engineered overall symmetry and electronic properties through weak interlayer vdW coupling in both homostructures[8–15] and heterostructures[16-19]. However, this approach has an inherent dilemma: the weak interlayer coupling that allows for easy exfoliation and assembly is also the interaction expected to determine the new properties of the resulting stacked assembly.

One promising approach is to modify the electronic properties not through weak interlayer coupling, but by directly engineering the intralayer atomic structure and breaking the intrinsic symmetries of a 2D crystal. Janus TMDs[21,22] represent a prime example of this powerful strategy. By replacing the chalcogen atoms on only one side of a monolayer (transforming a symmetric $X$–$M$–$X$ structure into an asymmetric $X$–$M$–$Y$ structure) (Figure 1a), the out-of-plane mirror symmetry is broken. This atomic-level re-engineering is predicted to induce a host of remarkable properties absent in the corresponding symmetric monolayers, including a built-in out-of-plane dipole and Rashba-type spin splitting.[21–24] Janus monolayer TMDs such as MoSSe and WSSe have been successfully synthesized and their unique optical properties have been explored.[20,21,25–30] However, the instability and rapid degradation of Janus monolayers in ambient conditions,[29] combined with the difficulty of fabricating large-area monolayer Janus TMDs, have limited the available experimental probes, leaving their detailed electronic structure largely unexplored. Thus, a precise momentum-resolved investigation of electronic structure is crucial for a fundamental understanding of how atomic-level symmetry engineering manifests directly onto the electronic band dispersions that govern the material's novel properties.

In this study, we directly observe the evolution of the electronic band structure from monolayer $WSe_2$ to Janus WSSe on the same monolayer flake (Figure 1a) by using the micro-focused angle-resolved photoemission spectroscopy (μ-ARPES). The transformation from monolayer $WSe_2$ to monolayer Janus WSSe was achieved at room temperature by replacing the top-layer Se atoms with S atoms via an $H_2$-plasma-assisted chalcogen-exchange method.[25,30] We successfully elucidate band structure modifications arising from the sulfur substitution, revealing both the lifting of spin degeneracy due to the broken horizontal mirror symmetry and a significant change in the dispersion of the highest valence band, which governs the optoelectronic properties.

The monolayer WSe$_2$/graphite/h-BN van der Waals heterostructure on SiO$_2$/Si substrate was fabricated using an all-dry pick-up,[31,32] and flip method[33] using an Elvacite 2552C copolymer inside a N$_2$ glove box chamber.[31,32] Monolayer WSe$_2$ flakes were mechanically exfoliated from a chemical-vapor-transport–grown bulk WSe$_2$ single crystal (2D Semiconductors, Inc.). The h-BN, graphite and WSe$_2$ flakes were sequentially picked-up by a polymer stamp (Elvacite2552C). The assembled heterostructure was once transferred to another polymer stamp at room temperature to turn over the heterostructure face[33,34] and dropped onto a SiO$_2$/Si substrate with a prepatterned metal electrode as schematically shown in Figure 1b. The polymer residues were removed with chloroform. To suppress the sample charging caused by the photoemission process, the WSe$_2$ flake is electrically connected to ground via the stacked graphite and Au/Ti electrode as shown in Figure 1b,c. Conversion to Janus WSSe (see details in Supporting Information) was carried out by H$_2$ plasma treatment at room temperature [25,29,30] while the photoluminescence (PL) peak position was monitored *in situ* using a 532 nm laser[26,30], as schematically shown in Figure 1d. The fabrication details of the Janus WSSe monolayer are described elsewhere.[26,30] Figure 1e displays PL spectra acquired at 1 min. intervals during the H$_2$-plasma–assisted chalcogen exchange. The starting spectrum shows the 1.6 eV A-exciton emission of pristine monolayer WSe$_2$, which originates from the direct optical transition at K–point. As the H$_2$ plasma treatment proceeds, this feature shifts and evolves into a new peak near 1.8 eV. These spectral changes are consistent with previous reports of the optical response of monolayer WSe$_2$ and WSSe.[25,26,30] After 10 min of plasma treatment, the monolayer Janus WSSe was capped with amorphous sulfur to inhibit ambient degradation. The cap was subsequently desorbed by annealing in ultrahigh vacuum at 180 °C for 3 hours prior to μ-ARPES measurement. μ-ARPES measurements were performed at BL28 in the Photon Factory, KEK[35] using photon energy of 100 eV. During the measurement, the sample manipulator temperature was kept below 20 K. The first principles calculations were performed within the frame of density functional theory using the OpenMX code. [36–38] The Perdew–Burke–Ernzerhof (PBE) functional within the generalized gradient approximation (GGA)[39] was used to optimize the lattice parameter and atomic coordination of monolayer WSe$_2$ and WSSe. We also used numerically localized basis sets of W7.0-s3p3d2f1, Se7.0-s2p2d2f1, S7.0-s2p2d2f1 where *Xr*-s$n_s$p$n_p$d$n_d$f$n_f$ stands for a PAO basis set for element *X* consisted of the $n_s$ *s*-orbitals, $n_p$ *p*-orbitals, $n_d$ *d*-orbitals, and $n_f$ *f*-orbitals with the cutoff radius of r in units of Bohr. We used the norm-conserving pseudopotentials including the 3*s* and 3*p*, the 4*s* and 4*p*, and the 5*s*, 5*p*, and 5*d* electrons for sulfur, selenium, and tungsten as valence electrons, respectively. A cutoff energy value of 300 Ry was used for charge density. Non-collinear spin density functional with two-component spinor wavefunctions[40,41] was used to consider the electronic structures, and spin-orbit interaction was taken into account in a fully relativistic treatment of the total angular momentum-dependent pseudopotential[42]. Optimized lattice parameter *a* = 3.318 (3.254) Å and distance between W and chalcogen atoms $d_{W-Se}$ = 2.547 Å ($d_{W-Se}$ = 2.547 Å, $d_{W-S}$ = 2.437 Å) for monolayer WSe$_2$ (WSSe) are consistent with the previous calculational study. [43]

Figure 2a,b presents the ARPES images of pristine monolayer WSe$_2$ and Janus monolayer WSSe recorded along high-symmetry *M*–Γ–*K* momentum path shown in the inset of Figure 2a. For better visualization of the band dispersions, curvature plots[44] for the respective ARPES images are shown in Figure 2c,d. In the pristine sample (Figure 2a,c), we clearly observe the characteristic band dispersions of monolayer WSe$_2$, that the highest valence band lifts spin degeneracy along the Γ–*K* direction, forming the valence band maxima (VBM) at the Brillouin zone corner.[45–47] In contrast, along the Γ–*M* direction, the observed band dispersions forms spin degeneracy or negligible spin splitting. Those band configurations are clearly confirmed in the calculated band

dispersions in Figure 2e and explained by symmetrically allowed spin-momentum locking state in the point group $D_{3h}$ system of monolayer WSe$_2$.[48,49] The combination of three-fold rotational symmetry ($C_3$) and spin–orbit coupling forms with staggered out-of-plane spin polarizations, as schematically illustrated in the inset of Figure 2c. Along the Γ–M direction, the spin-degenerate band dispersions derive from the co-existence of the parallel and perpendicular mirror planes against the z-axis. In the current experimental condition, although the horizontal mirror plane is naively broken due to the van der Waals stacking structure, a graphite back-electrode exerts no discernible influence on the band dispersion of monolayer WSe$_2$, indicating that any substrate-induced symmetry breaking is below our detection limit.

Figure 2b shows the ARPES image from the same sample after the the H$_2$-plasma-assisted chalcogen exchange, capturing the electronic band structure of the resulting Janus monolayer WSSe. Although the ARPES spectrum of Janus monolayer WSSe seems to become broader than that of pristine monolayer WSe$_2$, the curvature plot in Figure 2d enables to resolve the individual band dispersions. Overall observed dispersion shapes are similar between monolayer WSe$_2$ and Janus WSSe. Indeed, our first-principles calculations also show similar overall band dispersions between them (Figure 2e,f). However, the calculational result of monolayer Janus WSSe (Figure 2f) reveals a qualitatively pronounced modification of the dispersion along the Γ–M direction. As denoted by the black arrow in Figure 2f, we can see spin-splitting of approximately 200 meV, which is absent in the monolayer WSe$_2$. This feature can be ascribed to Rashba-type spin–momentum locking[50] that emerges from the lack of the horizontal mirror plane in Janus WSSe as described in Figure 1a. Thus, the $C_{3v}$ symmetry of monolayer Janus WSSe allows for a complex spin texture combining both Rashba-type in-plane and staggered out-of-plane spin polarizations as schematically illustrated in the inset of Figure 2d. In the ARPES results in figure 2b,d, a faint but finite ARPES intensities consistent with the calculated Rashba-type spin splitting are observed at the region indicated by the white arrow in Figure 2b, whereas no such intensity is detected for pristine monolayer WSe$_2$. To enhance the visibility of weak ARPES intensity features, the boxed regions around the Γ–point in Figure 2b,d are depicted with different color scales. In the curvature plot of Figure 2d, a dispersive segment-like signature rather than from noise or a nondispersive defect state can be observed.

To discuss the ARPES signature of Janus monolayer WSSe along the Γ–M direction in detail, Figure 3a presents an enlarged view of the rectangular region shown in Figure 2d. Figure 3b,c shows the energy distribution curves (EDCs) and ARPES curvature profiles extracted from the ARPES image in Figure 2b and the ARPES curvature plot in Figure 2d, respectively. The corresponding measurement cuts from $k = -0.20$ Å$^{-1}$ to $-0.60$ Å$^{-1}$ at the interval of $0.05$ Å$^{-1}$ are represented by the white segments in Figure 3a. The peak positions of ARPES intensities are plotted with red and blue circle markers with error bars, estimated from cross-checking of the EDCs and curvature profiles. In addition to a dispersive band feature indicated by the blue circle markers in Figure 3a, a weaker dispersive branch appears on its low-binding-energy shoulder of the EDCs and can be traced between $k = -0.30$ Å$^{-1}$ and $-0.55$ Å$^{-1}$ (red circle markers). In the curvature profiles, peak structures in the ARPES intensity appear as dips. Indeed, the curvature profiles in Figure 3c clearly exhibit dips at the positions indicated by the red markers, corroborating the existence of this weaker dispersive branch. The band splitting at $k = -0.35$ Å$^{-1}$ is evaluated to be approximately 270 meV. This value is comparable to the splitting of ~200 meV obtained from our first-principles band calculations (Fig. 2f), indicating good consistency. These experimental features, observed only after the plasma treatment, are consistent with the theoretically expected band dispersion of Janus WSSe, thereby providing a direct observation of the band structure modification.

We now focus on the modification of the highest valence band, which directly governs the optoelectronic properties. Figure 4a,b present the enlarged ARPES curvature plot along the Γ–K direction for monolayer WSe$_2$ and WSSe respectively. The purple and red markers represent the peak positions estimated from EDCs as shown in Figure 4c,d, respectively. Those band dispersion shapes near the K−point are remarkably similar for both materials. This is because the VBM are predominantly composed of the in-plane W 5$d$ orbitals, so breaking the horizontal mirror plane does not substantially modify the K–valley spin-orbit coupling itself.[20,48,49] Consistent with this picture, recent ARPES study on epitaxial Janus MoSSe/Au(111) reported a K–valley spin–orbit splitting.[51] The VBMs at the K–point locate at a binding energy of $E_B$ = 0.82 eV for both pristine WSe$_2$ and Janus WSSe. Previous experimental studies have established the electronic band gap of monolayer WSe$_2$ to be approximately 2.1–2.2 eV evaluated by scanning tunneling spectroscopy measurement.[53,54] Therefore, the Fermi level in our pristine WSe$_2$ sample lies within the band gap. The consistent binding energy of the VBM at the K–point between monolayer WSe$_2$ and Janus WSSe can be attributed to the Fermi level pinning[47,55,56] at the interface with the graphite substrate. Since the bottom selenium layer remains in direct contact with the graphite even after the conversion to Janus WSSe, this pinning effect is likely to be preserved. Regarding to the Γ–point, the top of the valence band exhibits a significant upward shift of 0.16 eV from $E_B$ = 1.35 eV in pristine monolayer WSe$_2$ (Figure 4a,c) to $E_B$ = 1.19 eV in Janus monolayer WSSe (Figure 4b,d). The highest valence band at the Γ–point is formed by the hybridization of out-of-plane W 5$d_{z^2}$ and chalcogen $p_z$ orbitals,[24,49] resulting in a relatively larger contribution from the chalcogen atoms compared to the VBM at the K–point. The result of the first principles calculation in this study well reproduces with a comparable energy shift of approximately 0.22 eV at the Γ-point (Figure 2e,f). Based on this calculational result, we evaluate the orbital composition ratio for the highest valence band at the Γ–point.[57] For the Janus WSSe monolayer, this yields a projected orbital composition ratio of W 5$d_{z^2}$:Se4$p_z$:S3$p_z$ ≈ 12:1:4, which is qualitatively consistent with previous calculational study.[24] The contribution from the S3$p_z$ orbital, four times larger than that of the Se 4$p_z$ orbital, suggests that the electronic structure of the Janus monolayer at the Γ–point is more akin to that of monolayer WS$_2$ than WSe$_2$. For a direct comparison, we also performed first-principles calculations for a free-standing monolayer WS$_2$ with optimized crystal structure ($a$ = 3.192 Å, $d_{W-S}$ = 2.429 Å). Figure 4e summarizes the energy positions of the highest valence band at the Γ–point relative to the respective VBM at the K–point for monolayer WSe$_2$, Janus WSSe, and WS$_2$. The circles show the results from ARPES; for WS$_2$, the value was extracted by referencing the ARPES results of Henck et al.[58] The black bars show the results of the first-principles band calculations in this study. Indeed, as expected from the orbital composition obtained from our first-principles calculations, the energy position of the Γ–point in Janus WSSe is closer to that of WS$_2$ than to WSe$_2$. This trend is even more pronounced in ARPES result. These direct observations experimentally confirm the crucial role of chalcogen orbital character in determining the band dispersions of Janus TMDs.

In conclusion, we have directly observed the electronic band structure evolution from monolayer WSe$_2$ to Janus WSSe using μ-ARPES on an identical sample. In monolayer Janus WSSe, we observe the ARPES signatures consistent with the consequences of broken horizontal mirror symmetry, notably the emergence of a Rashba-type spin splitting along the Γ–M direction. Our momentum-resolved analysis reveals that while the VBM at the K–point retains its dispersive shape and remains pinned by the graphite substrate, the band at the Γ–point shifts significantly, a behavior governed by its evolving chalcogen orbital character. This work demonstrates a powerful approach to tailoring electronic structures of Janus monolayer TMDs, opening avenues for the design of future valleytronic and optoelectronic devices.

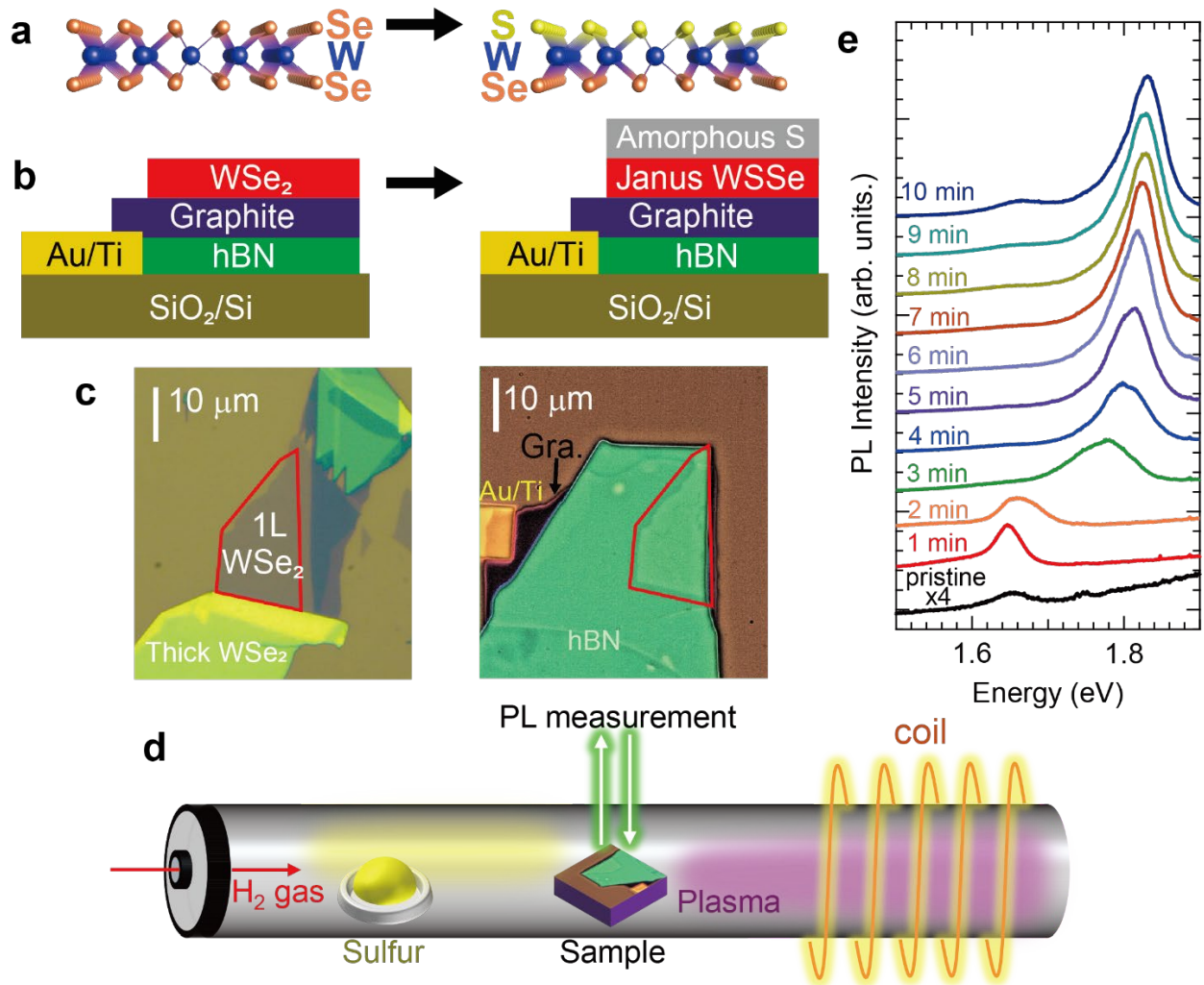

**Figure 1.** Sample fabrication of monolayer $WSe_2$ and Janus WSSe for μ-ARPES measurements. (a) Crystal structures of monolayer $WSe_2$ (left) and Janus WSSe (right). (b) Schematics of the van der Waals stacked monolayer $WSe_2$ (left) and Janus WSSe (right) samples for μ-ARPES measurements. An additional amorphous-sulfur cap is introduced in the Janus WSSe sample to suppress ambient degradation. The sulfur cap is removed before μ-ARPES measurement by annealing (~180 °C, 3 hours) in ultrahigh vacuum chamber. (c) Optical microscope images of the used monolayer $WSe_2$ flake (left) and fabricated monolayer $WSe_2$/graphite/h-BN heterostructure sample (right). (d) Schematic of the experimental setup for the conversion process from monolayer $WSe_2$ to Janus WSSe using $H_2$ plasma treatment and *in situ* photoluminescence (PL) measurement. (e) *In situ* PL spectra collected at 1 min intervals during the $H_2$ plasma treatment.

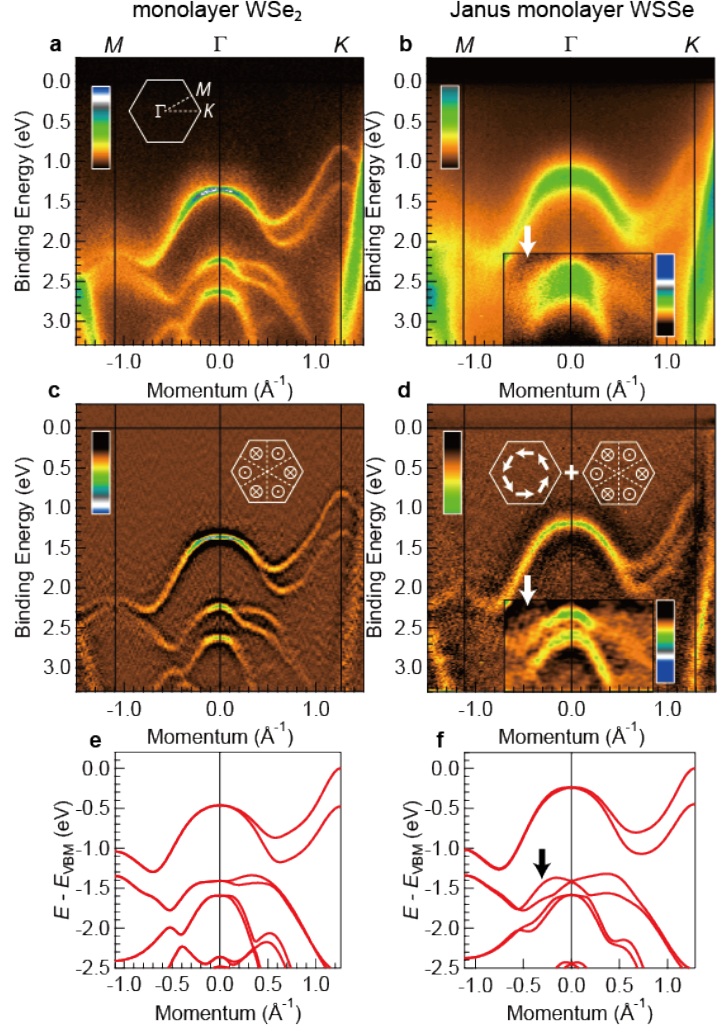

**Figure 2.** Electronic band structures of monolayer WSe2 and Janus WSSe. (a),(b) ARPES images recorded along $M$–$\Gamma$–$K$, indicated in the Brillouin zone inset of (a), for pristine monolayer WSe2 (a) and Janus WSSe (b). To highlight weak ARPES intensity features, the lower-energy portion of (b) is displayed with an independent color scale, indicated by the adjacent color bar. In (b), the white arrow indicates the faint but finite photoemission intensity that appears off the $\Gamma$-point after sulfur substitution. (c),(d) Images obtained by the curvature analysis,[44] which sharpen dispersive features, for the ARPES images in (a),(b), respectively. The lower-energy boxed region in (d) uses the alternative color scale and again represents the additional band feature (white arrow). The white insets in (c) and (d) schematically illustrate the symmetry-allowed spin polarization directions. For monolayer WSe2 with $D_{3h}$ symmetry (c), only the out-of-plane spin component is allowed. For Janus WSSe (d), the broken horizontal mirror symmetry ($C_{3v}$) additionally allows for an in-plane Rashba-type spin component. (e), (f) Calculated band dispersions obtained from fully relativistic density-functional theory (DFT) calculations including spin-orbit coupling for monolayer WSe2 (e) and Janus WSSe (f). In (f), the black arrow highlights the characteristic lifting of spin degeneracy in monolayer Janus WSSe, realized by the interplay of strong spin–orbit interaction and the broken horizontal-mirror symmetry.

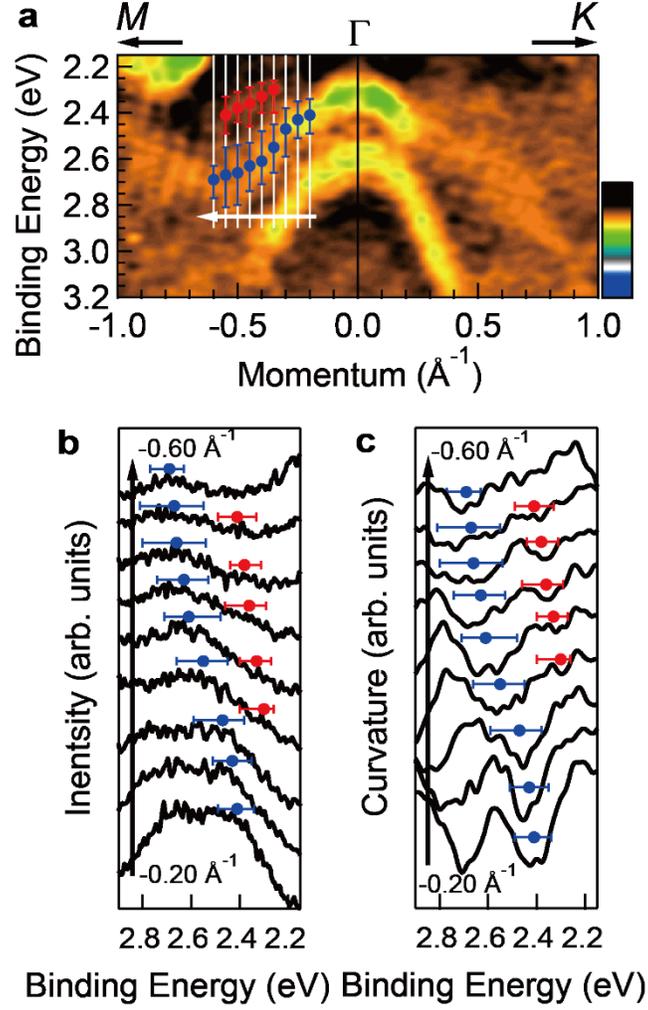

**Figure 3.** Lifting of spin degeneracy along Γ–M direction in monolayer Janus WSSe. (a) Magnified curvature plots of the boxed region in Figure 2d. White lines represent the measurement cuts for corresponding energy distribution curves (EDCs) shown in (b) and curvature profiles shown in (c). The blue and red circles with error bars are estimated peak position from both the EDCs in (b) and curvature profiles in (c). (b) EDCs extracted from the corresponding ARPES image shown in Figure 2b for the measurement cuts from –0.20 to –0.60 Å$^{-1}$ as shown in (a). (c) Curvature profiles obtained from (b).

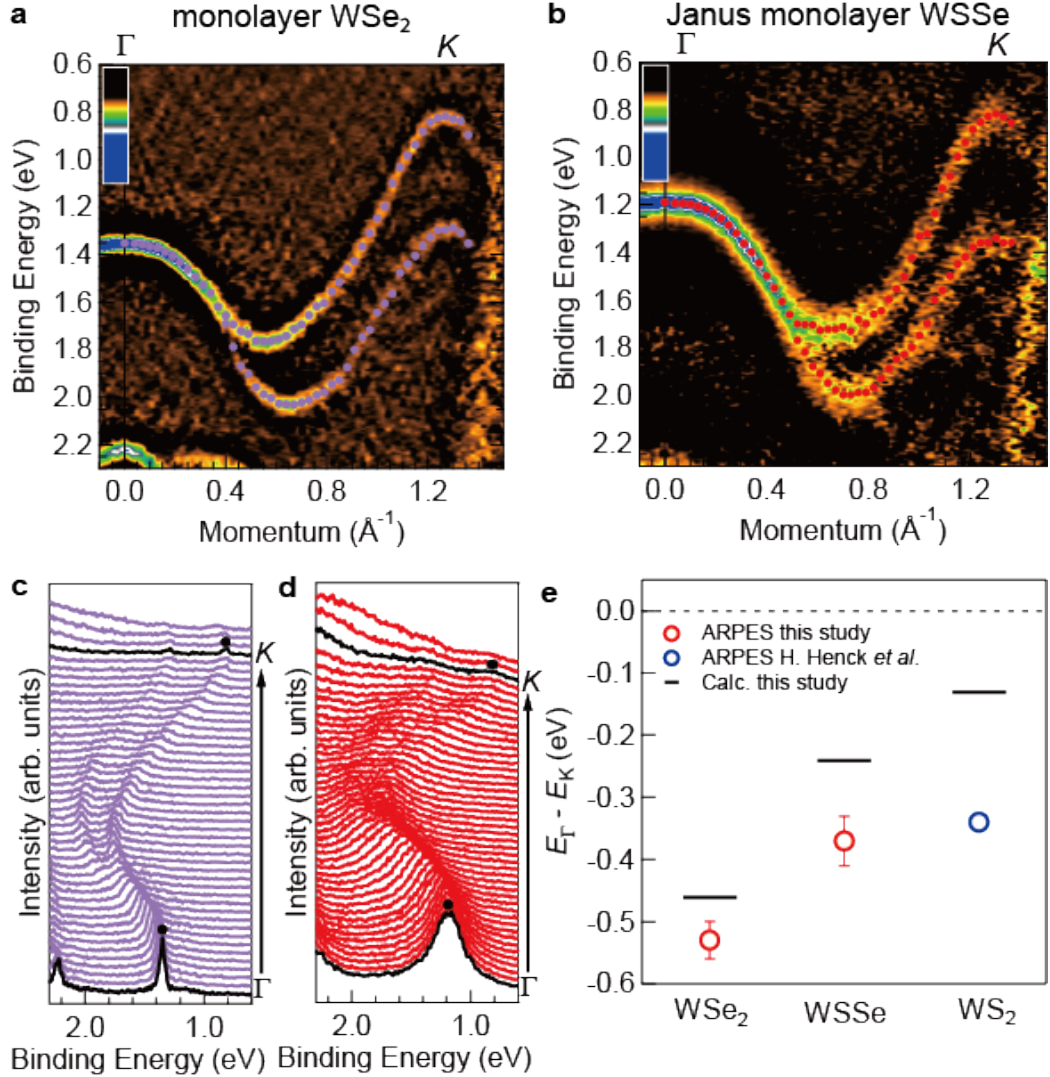

**Figure 4.** Modification of the highest valence band from monolayer WSe$_2$ to Janus WSSe. (a), (b) Curvature plots of the ARPES images for monolayer WSe$_2$ (a) and Janus WSSe (b). Purple and red circles represent peak positions estimated from EDCs shown in (c) and (d). (c), (d) EDCs extracted from the ARPES images in Figure 2a,b. The EDCs at high symmetrical Γ– and K–point and peak position of the highest valence band are represented by black color and markers. (e) Energy positions of the highest valence band at the Γ-point relative to the VBM at the K–point for monolayer WSe$_2$, Janus WSSe, and WS$_2$. The red circles with error bars represent the values estimated from our ARPES measurements in this study. The blue circle is a value taken from the ARPES study of Henck, H. *et al.*, *Phys. Rev. B* **2018**, *97*, 155421.[58] The black bars indicate the results obtained from our first-principles band calculations.

AUTHOR INFORMATION

**Corresponding Author**

*sakano@uec.ac.jp

**Present Addresses**

†Graduate School of Informatics and Engineering, The University of Electro-Communications, Chofu, Tokyo 182-8585, Japan

‡Department of Physics, Nagoya University, Nagoya, Aichi 464-8602, Japan

**Author Contributions**

K.W. and T.T. synthesized *h*-BN single crystals. WSe$_2$/graphite/*h*-BN samples were fabricated by S. A., and T.Y. with help from M. S., Y. S., M. S., and T. M. H$_2$ plasma treatment and PL measurement was carried out by M. S, S. A, T. Y, S. T., B. D., O. H., and T. K. ARPES measurements were carried out by M. S. and S. A. with help from N. M., M. K., K. H., K. O., K. S., S. S., and T. S. M. S. and S. A. analyzed ARPES data with help from N. M. N. Y. and F. I. performed DFT calculations. M.S. and K. I. designed the project and wrote the manuscript with input from the other authors.

**Funding Sources**

This work was partly supported by a CREST project (Grants No. JPMJCR20B4, JPMJCR23A2) from the Japan Science and Technology Agency (JST), Japan Society for the Promotion of Science KAKENHI (Grants-in-Aid for Scientific Research) (Grants No. JP21H05232, JP21H05233, JP21H05234, JP21H05235, JP22H05452, JP23H00097, JP23K17756, JP24H01165, JP24H01166, JP24K01285, JP25H01525) and JST PRESTO (Grant No. JPMJPR20A8).

**Notes**
ACKNOWLEDGMENT
This work was partly performed under the approval of the Photon Factory Program Advisory Committee (Proposal No. 2023G088, 2023G108, 2024G002, 2024G128, and No. 2021S2-001). The computation in this work has been done using the facilities of the Supercomputer Center, the Institute for Solid State Physics, the University of Tokyo.

REFERENCES

(1) Novoselov, K. S.; Geim, A. K.; Morozov, S. V.; Jiang, D.; Zhang, Y.; Dubonos, S. V.; Grigorieva, I. V.; Firsov, A. A. Electric Field Effect in Atomically Thin Carbon Films. *Science* **2004**, *306*, 666–669.

(2) Watanabe, K.; Taniguchi, T.; Kanda, H. Direct-bandgap properties and evidence for ultraviolet lasing of hexagonal boron nitride single crystal. *Nat. Mater.* **2004**, *3*, 404–409.

(3) Dean, C. R.; Young, A. F.; Meric, I.; Lee, C.; Wang, L.; Sorgenfrei, S.; Watanabe, K.; Taniguchi, T.; Kim, P.; Shepard, K. L.; Hone, J. Boron nitride substrates for high-quality graphene electronics. *Nat. Nanotechnol.* **2010**, *5*, 722–726.

(4) Mak, K. F.; Lee, C.; Hone, J.; Shan, J.; Heinz, T. F. Atomically thin $MoS_2$: A new direct-gap semiconductor. *Phys. Rev. Lett.* **2010**, *105*, 136805.

(5) Splendiani, A.; Sun, L.; Zhang, Y.; Li, T.; Kim, J.; Chim, C.-Y.; Galli, G.; Wang, F. Emerging photoluminescence in monolayer $MoS_2$. *Nano Lett.* **2010**, *10*, 1271–1275.

(6) Li, L.; Yu, Y.; Ye, G. J.; Ge, Q.; Ou, X.; Wu, H.; Feng, D.; Chen, X. H.; Zhang, Y. Black Phosphorus Field-Effect Transistors. *Nat. Nanotechnol.* **2014**, *9*, 372–377.

(7) Geim, A. K.; Grigorieva, I. V. Van der Waals heterostructures. *Nature* **2013**, *499*, 419–425.

(8) Bistritzer, R.; MacDonald, A. H. Moiré Bands in Twisted Double-Layer Graphene. *Proc. Natl. Acad. Sci. U.S.A.* **2011**, *108*, 12233–12237.

(9) Cao, Y.; Fatemi, V.; Fang, S.; Watanabe, K.; Taniguchi, T.; Kaxiras, E.; Jarillo-Herrero, P. Unconventional Superconductivity in Magic-Angle Graphene Superlattices. *Nature* **2018**, *556*, 43–50.

(10) Yasuda, K.; Wang, X.; Watanabe, K.; Taniguchi, T.; Jarillo-Herrero, P. Stacking-Engineered Ferroelectricity in Bilayer Boron Nitride. *Science* **2021**, *372*, 1458–1462.

(11) Wang, X.; Yasuda, K.; Zhang, Y.; Liu, S.; Watanabe, K.; Taniguchi, T.; Hone, J.; Fu, L.; Jarillo-Herrero, P. Interfacial ferroelectricity in rhombohedral-stacked bilayer transition metal dichalcogenides. *Nat. Nanotechnol.* **2022**, *17*, 367–371.

(12) Cai, J.; Anderson, E.; Wang, C.; Zhang, X.; Liu, X.; Holtzmann, W.; Zhang, Y.; Fan, F.; Taniguchi, T.; Watanabe, K.; Ran, Y.; Cao, T.; Fu, L.; Xiao, D.; Yao, W.; Xu, X. Signatures of fractional quantum anomalous Hall states in twisted $MoTe_2$. *Nature* **2023**, *622*, 63–68.

(13) Zeng, Y.; Xia, Z.; Kang, K.; Zhu, J.; Knüppel, P.; Vaswani, C.; Watanabe, K.; Taniguchi, T.; Mak, K. F.; Shan, J. Thermodynamic evidence of fractional Chern insulator in moiré $MoTe_2$. *Nature* **2023**, *622*, 69–73.

(14) Park, H.; Cai, J.; Anderson, E.; Zhang, Y.; Zhu, J.; Liu, X.; Wang, C.; Holtzmann, W.; Hu, C.; Liu, Z.; Taniguchi, T.; Watanabe, K.; Chu, J.-H.; Cao, T.; Fu, L.; Yao, W.; Chang, C.-Z.; Cobden, D.; Xiao, D.; Xu, X. Observation of fractionally quantized anomalous Hall effect. *Nature* **2023**, *622*, 74–79.

(15) Xu, F.; Sun, Z.; Jia, T.; Liu, C.; Xu, C.; Li, C.; Gu, Y.; Watanabe, K.; Taniguchi, T.; Tong, B.; Jia, J.; Shi, Z.; Jiang, S.; Zhang, Y.; Liu, X.; Li, T. Observation of integer and fractional quantum anomalous Hall effects in twisted bilayer $MoTe_2$. *Phys. Rev. X* **2023**, *13*, 031037.

(16) Tang, Y.; Li, L.; Li, T.; Xu, Y.; Liu, S.; Barmak, K.; Watanabe, K.; Taniguchi, T.; MacDonald, A. H.; Shan, J.; Mak, K. F. Simulation of Hubbard Model Physics in $WSe_2/WS_2$ Moiré Superlattices. *Nature* **2020**, *579*, 353–358.

(17) Regan, E. C.; Wang, D.; Jin, C.; Utama, M. I. B.; Gao, B.; Wei, X.; Zhao, S.; Zhao, W.; Zhang, Z.; Yumigeta, K.; Blei, M.; Carlström, J. D.; Watanabe, K.; Taniguchi, T.; Tongay, S.; Crommie, M.; Zettl, A.; Wang, F. Mott and Generalized Wigner Crystal States in $WSe_2/WS_2$ Moiré Superlattices. *Nature* **2020**, *579*, 359–363

(18) Akamatsu, T.; Ideue, T.; Zhou, L.; Dong, Y.; Kitamura, S.; Yoshii, M.; Yang, D.; Onga, M.; Nakagawa, Y.; Watanabe, K.; Taniguchi, T.; Laurienzo, J.; Huang, J.; Ye, Z.; Morimoto, T.; Yuan, H.; Iwasa, Y. A van der Waals interface that creates in-plane polarization and a spontaneous photovoltaic effect. *Science* **2021**, *372*, 68–72.

(19) Rogée, L.; Wang, L.; Zhang, Y.; Cai, S.; Wang, P.; Chhowalla, M.; Ji, W.; Lau, S. P. Ferroelectricity in untwisted heterobilayers of transition metal dichalcogenides. *Science* **2022**, *376*, 973–978.

(20) Lu, A.-Y.; Zhu, H.; Xiao, J.; Chuu, C.-P.; Han, Y.; Chiu, M.-H.; Cheng, C.-C.; Yang, C.-W.; Wei, K.-H.; Yang, Y.; Wang, Y.; Sokaras, D.; Nordlund, D.; Yang, P.; Muller, D. A.; Chou, M.-Y.; Zhang, X.; Li, L.-J. Janus monolayers of transition metal dichalcogenides. *Nat. Nanotechnol.* **2017,** *12*, 744–749.

(21) Zhang, J.; Jia, S.; Kholmanov, I.; Dong, L.; Er, D.; Chen, W.; Guo, H.; Jin, Z.; Shenoy, V. B.; Shi, L.; Lou, J. Janus monolayer transition-metal dichalcogenides. *ACS Nano* **2017**, *11*, 8192–8198.

(22) Hu, T.; Jia, F.; Zhao, G.; Wu, J.; Stroppa, A.; Ren, W. Intrinsic and anisotropic Rashba spin splitting in Janus transition-metal dichalcogenide monolayers *Phys. Rev. B* **2018**, *97*, 235404.

(23) Xia, C.; Du, J.; Li, M.; Areshkin, D. A.; Wang, T.; Wei, S.-H.; Li, H. Universality of electronic characteristics and photocatalyst applications in the two-dimensional Janus transition metal dichalcogenides *Phys. Rev. B* **2018**, *98*, 165424.

(24) Zhou, W.; Chen, J.; Yang, Z.; Liu, J.; Ouyang, F. Geometry and Electronic Structure of Monolayer, Bilayer, and Multilayer Janus WSSe. *Phys. Rev. B* **2019**, *99*, 075160.

(25) Trivedi, D. B.; Turgut, G.; Qin, Y.; Sayyad, M. Y.; Hajra, D.; Howell, M.; Liu, L.; Yang, S.; Patoary, N. H.; Li, H.; Petrić, M. M.; Meyer, M.; Kremser, M.; Barbone, M.; Soavi, G.; Stier, A. V.; Müller, K.; Yang, S.; Esqueda, I. S.; Zhuang, H.; Finley, J. J.; Tongay, S. Room-temperature synthesis of 2D Janus crystals and their heterostructures. *Adv. Mater.* **2020**, *32*, e2006320.

(26) Kaneda, M.; Zhang, W.; Liu, Z.; Gao, Y.; Maruyama, M.; Nakanishi, Y.; Nakajo, H.; Aoki, S.; Honda, K.; Ogawa, T.; Hashimoto, K.; Endo, T.; Aso, K.; Chen, T.; Oshima, Y.; Yamada-Takamura, Y.; Takahashi, Y.; Okada, S.; Kato, T.; Miyata, Y. Nanoscrolls of Janus monolayer transition metal dichalcogenides. *ACS Nano* **2024**, *18*, 2772–2781.

(27) Zheng, T.; Lin, Y.-C.; Yu, Y.; Valencia-Acuna, P.; Puretzky, A. A.; Torsi, R.; Liu, C.; Ivanov, I. N.; Duscher, G.; Geohegan, D. B.; Ni, Z.; Xiao, K.; Zhao, H. Excitonic Dynamics in Janus MoSSe and WSSe Monolayers. *Nano Lett.* **2021**, *21*, 931–937.

(28) Petrić, M. M.; Kremser, M.; Barbone, M.; Qin, Y.; Sayyad, M. Y.; Shen, Y.; Tongay, S.; Finley, J. J.; Müller, K. Raman spectrum of Janus transition metal dichalcogenide monolayers WSSe and MoSSe. *Phys. Rev. B* **2021**, *103*, 035414.

(29) Bai, Y.; Maity, I.; Zhu, W.; Mu, R.; Li, K.; Zhou, J.; Tongay, S.; Tan, S. J. R.; Gao, W. L.; Eda, G. Identification of exciton complexes in charge-tunable Janus WSeS monolayers. *ACS Nano* **2023**, *17*, 7326–7334.

(30) Zhang, W.; Liu, Z.; Nakajo, H.; Aoki, S.; Wang, H.; Wang, Y.; Gao, Y.; Maruyama, M.; Kawakami, T.; Makino, Y.; Kaneda, M.; Chen, T.; Aso, K.; Ogawa, T.; Endo, T.; Nakanishi, Y.; Watanabe, K.; Taniguchi, T.; Oshima, Y.; Yamada-Takamura, Y.; Koshino, M.; Okada, S.; Matsuda, K.; Kato, T.; Miyata, Y. Chemically tailored semiconductor moiré superlattices of Janus heterobilayers. *Small Structures* **2024**, *5*, 2300514.

(31) Masubuchi, S.; Morimoto, M.; Morikawa, S.; Onodera, M.; Asakawa, Y.; Watanabe, K.; Taniguchi, T.; Machida, T. Autonomous robotic searching and assembly of two-dimensional crystals to build van der Waals superlattices. *Nat. Commun.* **2018**, *9*, 1413.

(32) Masubuchi, S.; Watanabe, E.; Seo, Y.; Okazaki, S.; Sasagawa, T.; Watanabe, K.; Taniguchi, T.; Machida, T. Deep-learning-based image segmentation integrated with optical microscopy for automatically searching for two-dimensional materials. *npj 2D Mater. Appl.* **2020**, *4*, 3.

(33) Masubuchi, S.; Sakano, M.; Tanaka, Y.; Wakafuji, Y.; Yamamoto, T.; Okazaki, S.; Watanabe, K.; Taniguchi, T.; Li, J.; Ejima, H.; Sasagawa, T.; Ishizaka, K.; Machida, T. Dry pick-and-flip assembly of van der Waals heterostructures for microfocus angle-resolved photoemission spectroscopy. *Sci. Rep.* **2022**, *12*, 10936.

(34) Sakano, M.; Tanaka, Y.; Masubuchi, S.; Okazaki, S.; Nomoto, T.; Oshima, A.; Watanabe, K.; Taniguchi, T.; Arita, R.; Sasagawa, T.; Machida, T.; Ishizaka, K. Odd-even layer-number effect of valence-band spin splitting in $WTe_2$. *Phys. Rev. Research* **2022**, *4*, 023247.

(35) Kitamura, M.; Souma, S.; Honma, A.; Wakabayashi, D.; Tanaka, H.; Toyoshima, A.; Amemiya, K.; Kawakami, T.; Sugawara, K.; Nakayama, K.; Yoshimatsu, K.; Kumigashira, H.; Sato, T.; Horiba, K. Development of a versatile micro-focused angle-resolved photoemission spectroscopy system with Kirkpatrick–Baez mirror optics. *Rev. Sci. Instrum.* **2022**, *93*, 033906.

(36) Ozaki, T. Variationally optimized atomic orbitals for large-scale electronic structures. *Phys. Rev. B* **2003**, *67*, 155108.

(37) Ozaki, T.; Kino, H. Numerical atomic basis orbitals from H to Kr. *Phys. Rev. B* **2004**, *69*, 195113.

(38) Ozaki, T.; Kino, H. Efficient projector expansion for the ab initio LCAO method. *Phys. Rev. B* **2005**, *72*, 045121.

(39) Perdew, J. P.; Burke, K.; Ernzerhof, M. Generalized gradient approximation made simple. *Phys. Rev. Lett.* **1996**, *77*, 3865–3868.

(40) Barth, U. von; Hedin, L. A local exchange-correlation potential for the spin polarized case. i. *J. Phys. C Solid State Phys.* **1972**, *5*, 1629–1642.

(41) Kubler, J.; Hock, K.-H.; Sticht, J.; Williams, A. R. Density functional theory of non-collinear magnetism. *J. Phys. F Met. Phys.* **1988**, *18*, 469–483.

(42) Theurich, G.; Hill, N. A. Self-consistent treatment of spin-orbit coupling in solids using relativistic fully separable ab initio pseudopotentials. *Phys. Rev. B* **2001**, *64*, 073106.

(43) Chaurasiya, R.; Dixit, A.; Pandey, R. Strain-mediated stability and electronic properties of WS2, Janus WSSe and WSe2 monolayers. *Superlattices Microstruct.* **2018**, *122*, 268–279.

(44) Zhang, P.; Richard, P.; Qian, T.; Xu, Y.-M.; Dai, X.; Ding, H. A precise method for visualizing dispersive features in image plots. *Rev. Sci. Instrum.* **2011**, *82*, 043712.

(45) Zhang, Y.; Ugeda, M. M.; Jin, C.; Shi, S.-F.; Bradley, A. J.; Martin-Recio, A.; Ryu, H.; Kim, J.; Tang, S.; Kim, Y.; Zhou, B.; Hwang, C.; Chen, Y.; Wang, F.; Crommie, M. F.; Hussain, Z.; Shen, Z.-X.; Mo, S.-K. Electronic Structure, Surface Doping, and Optical Response in Epitaxial $WSe_2$ Thin Films. *Nano Lett.* **2016**, *16*, 2485–2491.

(46) Wilson, N. R.; Nguyen, P. V.; Seyler, K.; Rivera, P.; Marsden, A. J.; Laker, Z. P. L.; Constantinescu, G. C.; Kandyba, V.; Barinov, A.; Hine, N. D. M.; Xu, X.; Cobden, D. H. Determination of Band Offsets, Hybridization, and Exciton Binding in 2D Semiconductor Heterostructures. *Sci. Adv.* **2017**, *3*, e1601832.

(47) Nguyen, P. V.; Teutsch, N. C.; Wilson, N. P.; Kahn, J.; Xia, X.; Graham, A. J.; Kandyba, V.; Giampietri, A.; Barinov, A.; Constantinescu, G. C.; Yeung, N.; Hine, N. D. M.; Xu, X.; Cobden, D. H.; Wilson, N. R. Visualizing Electrostatic Gating Effects in Two-Dimensional Heterostructures. *Nature* **2019**, *572*, 220–223.

(48) Xiao, D.; Liu, G.-B.; Feng, W.; Xu, X.; Yao, W. Coupled spin and valley physics in monolayers of MoS$_2$ and other group-VI dichalcogenides. *Phys. Rev. Lett.* **2012**, *108*, 196802.

(49) Liu, G.-B.; Shan, W.-Y.; Yao, Y.; Yao, W.; Xiao, D. Three-band tight-binding model for monolayers of group-VIB transition metal dichalcogenides. *Phys. Rev. B* **2013**, *88*, 085433.

(50) Ishizaka, K.; Bahramy, M. S.; Murakawa, H.; Sakano, M.; Shimojima, T.; Sonobe, T.; Koizumi, K.; Shin, S.; Miyahara, H.; Kimura, A.; Miyamoto, K.; Okuda, T.; Namatame, H.; Taniguchi, M.; Arita, R.; Nagaosa, N.; Kobayashi, K.; Murakami, Y.; Kumai, R.; Kaneko, Y.; Onose, Y.; Tokura, Y. Giant Rashba-Type Spin Splitting in Bulk BiTeI. *Nat. Mater.* **2011**, *10*, 521–526.

(51) Chaurasiya, R.; Dixit, A.; Pandey, R. Strain-mediated stability and electronic properties of WS$_2$, Janus WSSe and WSe$_2$ monolayers. *Superlattices Microstruct.* **2018**, *122*, 268–279.

(52) Picker, J.; Ghorbani-Asl, M.; Schaal, M.; Kretschmer, S.; Otto, F.; Gruenewald, M.; Neumann, C.; Fritz, T.; Krasheninnikov, A. V.; Turchanin, A. Atomic Structure and Electronic Properties of Janus SeMoS Monolayers on Au(111). *Nano Lett.* **2025**, *25*, 3330–3336.

(53) Zhang, C.; Chen, Y.; Johnson, A.; Li, M.-Y.; Li, L.-J.; Mende, P. C.; Feenstra, R. M.; Shih, C.-K. Probing Critical Point Energies of Transition Metal Dichalcogenides: Surprising Indirect Gap of Single Layer WSe$_2$. *Nano Lett.* **2015**, *15*, 6494–6500.

(54) Yankowitz, M.; McKenzie, D.; LeRoy, B. J. Local Spectroscopic Characterization of Spin and Layer Polarization in WSe$_2$. *Phys. Rev. Lett.* **2015**, *115*, 136803.

(55) Kang, J.; Liu, W.; Sarkar, D.; Jena, D.; Banerjee, K. Computational Study of Metal Contacts to Monolayer Transition-Metal Dichalcogenide Semiconductors. *Phys. Rev. X* **2014**, *4*, 031005.

(56) Allain, A.; Kang, J.; Banerjee, K.; Kis, A. *Electrical Contacts to Two-Dimensional Semiconductors. Nat. Mater.* **2015**, *14*, 1195–1205

(57) Mostofi, A. A.; Yates, J. R.; Lee, Y.-S.; Souza, I.; Vanderbilt, D.; Marzari, N. Wannier90: A tool for obtaining maximally-localised Wannier functions. *Comput. Phys. Commun.* **2008**, *178*, 685–699.

(58) Henck, H.; Ben Aziza, Z.; Pierucci, D.; Laourine, F.; Reale, F.; Palczynski, P.; Chaste, J.; Silly, M. G.; Bertran, F.; Le Fèvre, P.; Lhuillier, E.; Wakamura, T.; Mattevi, C.; Rault, J. E.; Calandra, M.; Ouerghi, *Phys. Rev. B* **2018**, *97*, 155421.